\documentclass[12pt]{article}
\usepackage{graphicx}
\usepackage{latexsym,amsmath,amsfonts,amsthm,amssymb}

\usepackage{times}

\usepackage[breaklinks,pdfpagemode=None,pdfstartview=FitH,pdfpagelayout=OneColumn]{hyperref}

\usepackage{graphics}

\title{Dimensional Reduction for Directed Branched Polymers}
\author{John Z. Imbrie\\
Department of Mathematics\\
University of Virginia\\
Charlottesville, VA  22904-4137\\}
\date{}
\begin{document}
\maketitle

\begin{abstract}
Dimensional reduction occurs when the critical behavior of one system can be related to that of another system in a lower dimension. We show that this occurs for directed branched polymers (DBP) by giving an exact relationship between DBP models in $D+1$ dimensions and repulsive gases at negative activity in $D$ dimensions.  This implies relations between exponents of the two models: $\gamma(D+1)=\alpha(D)$ (the exponent describing the singularity of the pressure), and $\nu_{\perp}(D+1)=\nu(D)$ (the correlation length exponent of the repulsive gas). It also leads to the relation $\theta(D+1)=1+\sigma(D)$, where $\sigma(D)$ is the Yang-Lee edge exponent.  We derive exact expressions for the number of DBP of size $N$ in two dimensions.

\vspace{7mm}
\end{abstract}

\newcommand{\dis}{\displaystyle}
\newcommand{\rarrow}{\rightarrow}

\newtheorem{theorem}{Theorem}
\newtheorem{definition}[theorem]{Definition}
\newtheorem{lemma}[theorem]{Lemma}
\newtheorem{example}[theorem]{Example}
\newtheorem{corollary}[theorem]{Corollary}
\newtheorem{proposition}[theorem]{Proposition}

\noindent
The phenomenon of dimensional reduction has attracted considerable attention over the years. The first example was the controversial random field Ising model (RFIM), whose critical behavior was conjectured to be the same as the pure Ising model in two fewer dimensions \cite{PS79}.  A proof of long-range order for the RFIM in three dimensions \cite{I84,I85} showed that dimensional reduction fails there, and recent work \cite{BDD98,F02,PS02} has elucidated what goes wrong.  A second example is the Parisi-Sourlas reduction of branched polymers (BP) in $D+2$ dimensions to the Yang-Lee edge or $i\varphi^3$ field theory in $D$ dimensions \cite{PS81}.  This was recently confirmed with the discovery of an exact relationship between BP models and repulsive gases at negative activity in two fewer dimensions \cite{BI01,BI03}.  The failure of the heuristic arguments for dimensional reduction in the RFIM underscores the importance of having an exact result. In this letter we give a third example, in which directed branched polymers (DBP) reduce to the repulsive gas at negative activity in one fewer dimension.

\begin{figure}
\begin{center}
\includegraphics[width=.6\columnwidth]{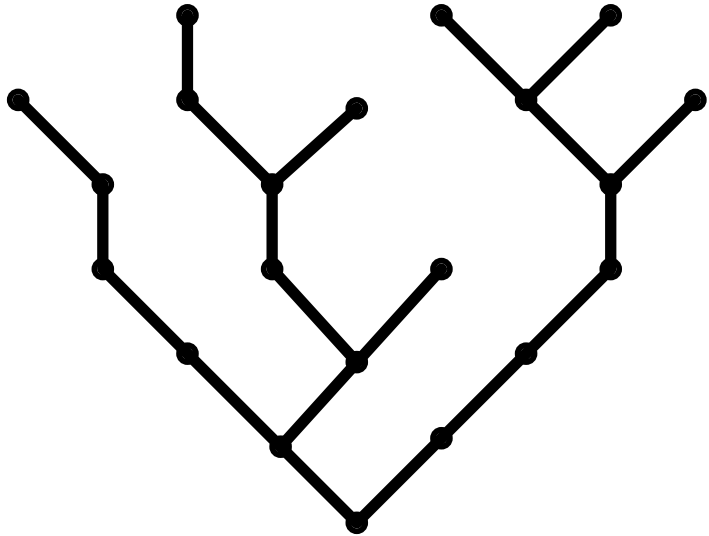}
\caption{\label{tree} A directed branched polymer in $\mathbb{Z}^2$.}
\end{center}
\end{figure}

We consider directed branched polymers as self-avoiding tree graphs embedded in $\mathbb{Z}^{D+1}$ or $\mathbb{R}^{D+1}$ so that every vertex can be reached from the root at $0$ by a sequence of links which move forward with respect to a preferred direction.  See Figure~\ref{tree}.  Let $d_N$ denote the number of DBP with $N$ vertices, and let $Z_{\mathrm{DBP}}(z)=\sum_N d_Nz^N$.  We prove the identity
\begin{equation}\label{eqn5}
\rho_{\mathrm{HC}}(z) = -Z_{\mathrm{DBP}}(-z),
\end{equation}
which relates the DBP generating function to the density of an associated repulsive gas model.  In contrast to the undirected case, the reduction in dimension is one.  If we define $\alpha$ from the singularity $\rho_{\mathrm{HC}} \sim (z-z_{\mathrm{c}})^{1-\alpha}$ of the hard-core gas at the negative activity critical point, and $\gamma$ from the singularity $Z_{\mathrm{DBP}} \sim (z-\tilde{z}_{\mathrm{c}})^{1-\gamma}$, then we find $\tilde{z}_{\mathrm{c}}=-z_{\mathrm{c}}$ and $\gamma(D+1)=\alpha(D)$. Define an exponent $\theta$ from the asymptotic behavior $d_N \sim z_c^{-N}N^{-\theta}$. Then $\theta=2-\gamma$.  Noting that the Yang-Lee edge exponent $\sigma$ \cite{F78} can be identified with $1-\alpha$ \cite{LF95,FP99}, we obtain the relation
\begin{equation}\label{eq1}
\theta(D+1)=1+\sigma(D). 
\end{equation}
In one and two dimensions, $\sigma$ can be determined from the exact values $\alpha(1)=\frac{3}{2}$ \cite{LF95}, $\alpha(2)=\frac{7}{6}$  (the latter follows from the solution to the hard-hexagon model \cite{B} at the negative activity critical point, see \cite{D83,BL87}; alternatively from \cite{C85}, assuming the model is in the Yang-Lee class.) Hence $\theta(2)=\frac{1}{2}$ and $\theta(3)=\frac{5}{6}$. We also obtain identities relating DBP correlations with repulsive gas correlations, which imply that the DBP exponent for the transverse correlation length $\nu_{\perp}(D+1)$ equals the repulsive gas exponent $\nu(D)$.

The related problem of directed lattice animals (DA) has been studied extensively \cite{C82,BJ82,SRY82,DL82,RY82,D82,D83,BL87,C01,D03,B98,RM01,K02}.  It is generally believed that the loop-free condition does not affect the critical behavior, so that both systems should have the same exponents.  For DA, there are exact results in two dimensions (see \cite{D82,C01,D03}, the review \cite{B98}, and references therein) and in three dimensions \cite{D83}. Also, in any dimension, models of DA have been related to dynamical models of hard-core lattice gases \cite{D83} (see also \cite{diF02}, which connects Lorentzian semi-random lattices with DA and gives an alternative derivation of Dhar's equivalence).  They have also been related to the critical dynamics of the Ising model in an imaginary field \cite{C82,BJ82}.  
As a result, the identity (\ref{eq1}) is believed to hold for DA \cite{C82,BJ82,SRY82}, and the values in $D=2, 3$ should be the same as the ones given above for DBP.

We consider a class of DBP models where the vertices $y_k = (t_k,x_k)$ have a time component $t_k \in \mathbb{R}_+$ and a space component $x_k \in S$, where $S$ is either $\mathbb{R}^D$ or $\mathbb{Z}^D$. If a lattice model is desired, the time component can be discretized by taking limits (see examples below).  Let $T$ be a tree graph on $\{1,\ldots,N\}$, and let $y_k$ be the position of the $k^{\mathrm{th}}$ vertex.
Fix the vertex 1 as the root, with $y_1 = 0$.  For each pair $(i,j)\equiv ij$, define
\begin{equation}
y_{ij} := (t_{ij},x_{ij}) := (|t_i-t_j|, \, x_i-x_j).
\end{equation}
Each link of $T$ connects a vertex $j$ to a vertex $i$, where $i$ is one step closer than $j$ to the root along $T$.
As we are considering {\em directed} BP, we require $t_j \geq t_i$. 

The weight associated with each DBP configuration depends on a linking weight $V(y) = V(t,x)$ and a repulsive weight $U(y) = U(t,x) = U(t,-x)$. The generating function for (rooted) DBP is written as
\begin{equation}\label{eqn1}
Z_{\mathrm{DBP}}(z) = \sum^\infty_{N=1} \frac{z^N}{(N-1)!} \sum_T \int_{(\mathbb{R}_{+}\times S)^{N-1}}
\prod_{ji \in T} [dy_{ji} V(y_{ji})] \prod_{ji \notin T} U(y_{ji})  .
\end{equation}
Here each pair $\{i,j\}$ appears exactly once, either in $\prod_{ji \in T}$ or in $\prod_{ji \notin T}$. We assume $U(t,x) \rarrow 1$ if either $t$ or $x$ tends to infinity (repulsion vanishes at infinite distance). In order to get dimensional reduction, we require
\begin{equation}\label{eqn2}
V(t,x) = U'(t,x),
\end{equation}
where prime denotes the $t$-derivative. For positive weights, we require that $U$ and $U' =V$ are positive. Also, $V$ needs to be an integrable function of $y$ for (\ref{eqn1}) to be well-defined.

Our main result relates $Z_{\mathrm{DBP}}$ to the density of a repulsive gas in the $D$-dimensional space $S$. Let $\Lambda \subset S$. Using the grand canonical partition function
\begin{equation}\label{eqn3}
Z_{\mathrm{HC}}(z) = \sum^\infty_{N=0} \frac{z^N}{N!} \int_{\Lambda^N} \prod^N_{i=1} dx_i \prod_{1 \leq i < j \leq N} U(0,x_{ij}),
\end{equation}
we can define the density
\begin{equation}\label{eqn4}
\rho_{\mathrm{HC}}(z) = z \, \frac{d}{dz} \left[\lim_{\Lambda \nearrow S} \frac{1}{|\Lambda|} \log Z_{\mathrm{HC}}(z)\right].
\end{equation}

\begin{theorem}\label{thm1}
For all $z$ such that the right-hand side converges absolutely,
\begin{equation}
\rho_{\mathrm{HC}}(z) = -Z_{\mathrm{DBP}}(-z).
\end{equation}
\end{theorem}

We now give some continuous and discrete examples of models of type (\ref{eqn1}).
If one takes $U(t,x) = \vartheta(t+|x|-1)$, where $\vartheta$ is the usual step function, then $V(t,x) = \delta(t+|x|-1)$. One obtains a DBP model where each monomer is a hard diamond (when $|x|= \sum^D_{\alpha=1}|x_\alpha|)$ or a hard double-cone (when $|x|^2 = \sum^D_{\alpha=1} x^2_\alpha$), obtained as $\{(t,x): |t|+|x| < 1\}$. Each monomer is distributed uniformly in contact with the positive surface of the monomer it is linked to (subject to the constraint of nonoverlap with other monomers). One can also consider a hard-sphere model by taking $U(t,x) = \vartheta(t^2+|x|^2-1)$.  Then the monomer is still distributed uniformly in $x$, which means angles close to the preferred direction are favored.

Theorem \ref{thm1} equates the generating functions of these $(D+1$)-dimensional DBP models to the density of the hard-sphere gas in $D$ dimensions, at negative activity. For $D=1$, the pressure of the hard-rod gas is computable; it is
\begin{equation}\label{eqn6}
p(z) = \mathtt{LambertW}(z) = -T(-z),
\end{equation}
where $T(z) = \sum^\infty_{N=1} z^N N^{N-1}/N!$ is the tree generating function \cite{BI03}. Hence
\begin{equation}\label{eqn7}
Z_{\mathrm{DBP}}(z) = -\rho_{\mathrm{HC}}(-z) = \sum^\infty_{N=1} \frac{z^NN^N}{N!}.
\end{equation}
This gives a simple expression for the volume available to DBP of size $N$; this can be checked with some effort for small values of $N$.

A natural class of lattice examples can be obtained by taking $U(t,x) = 1-I(x) \vartheta(1-t)$, where $I(x)$ is an indicator function of a set of ``neighbors'' in the lattice, such as $\{x: |x| \leq 1\}$.  Since $V(t,x) = I(x) \delta(t-1)$, the set determines which sites a link can jump to; $t$ always increases by 1. Figure~\ref{tree} shows a representative DBP for this model in two dimensions. On the gas side of identity (\ref{eqn5}), $t$ is set to 0, and we obtain various nearest-neighbor exclusion models, including the hard-hexagon model in $D=2$. 
When counting allowed DBP, it is important to include the factors $\prod_{ji \notin T} U(y_{ji})$, which enforce the nearest-neighbor exclusion for monomers on the same level (same value of $t$). Also, if a monomer at level $t$ has $n$ neighbors at level $t-1$ in the polymer, the associated $U$-factors can be written as $ \vartheta^{n-1}(0)$, which should be interpreted as $\frac{1}{n}=\int \vartheta^{n-1}d\vartheta$ (as can be seen by approximating $\vartheta$ with a smooth function and integrating over $t$). Monomers in levels more distant than this do not interact (similar conditions occur in Theorem III.2 of \cite{AR}; in fact (\ref{eqn10}) can be used to give an alternate proof of that identity).

Theorem \ref{thm1} equates the generating functions of these models with the nearest-neighbor exclusion models in $D$ dimensions associated with the weights $U(0,x) = 1-I(x)$. For the hard-hexagon model in $D=2$, there is an exact solution for the pressure \cite{B}. In $D=1$ we have a dimer model, for which the pressure is known, see \cite[Equation~2.16]{LF95}:
\begin{equation}\label{eqn8}
p(z) = \ln \left(\textstyle{\frac{1}{2}} + \textstyle{\frac{1}{2}} \sqrt{1+4z}\right).
\end{equation}
Thus the generating function is
\begin{equation}
Z_{\mathrm{DBP}}(z) = - z \, \frac{d}{dz} \, p(-z) = \sum^\infty_{N=1} \frac{[2N-1]!!2^{N-1}z^N}{N!},
\end{equation}
which enumerates the number of DBP with $N$ monomers.

In order to prove Theorem \ref{thm1} we derive an identity which encapsulates the effect of interpolating in the $t$ variables. Let $f(\mathbf{t})$ be a smooth function of $\{t_i\}, \{t_{ij}\}$ which approaches 0 when any of the $t$'s tends to infinity. Then a {\em forest-root formula} holds:
\begin{equation}\label{eqn10}
f(\mathbf{0}) = \sum_{(F,R)} \int_{\mathbb{R}^N_{+}}
\prod_{r \in R} [-dt_r] \prod_{ji \in F}[-d(t_j-t_i)] f^{(F,R)}(\mathbf{t}).
\end{equation}
Here $R$ is called the set of {\em roots} and is any subset of $\{1,\ldots,N\}$. We sum $F$ over the set of all {\em forests} or loop-free graphs on $\{1,\ldots,N\}$ with the property that each connected component or {\em tree} of $F$ contains exactly one root. Each link in a tree of $F$ connects a vertex $j$ to a vertex $i$ which is one step closer to the root for that tree. The integration region is $\{t_r \geq 0, \ r \in R \mbox{ and } t_j \geq t_i, \ ji \in F\}$. 

For $N=1$, (\ref{eqn10}) reduces to $f(0) = - \int^\infty_0 f'(t)dt$. For $N=2$, consider $f(t_1,t_2,t_{12})$ and use subscripts $1,2,12$ to denote partial derivatives. Then
\begin{equation}\label{eqn11}
f(\mathbf{0}) = - \int^\infty_0 ds (f_1(s,s,0) +f_2(s,s,0)).
\end{equation}
Apply the $N=1$ formula to the $f_1$ term, integrating with respect to $x_2-x_1$, and to the $f_2$ term, integrating with respect to $x_1-x_2$. The result is
\begin{equation}
f(\mathbf{0}) = \int^\infty_0 dt_1 \int^\infty_0 d(t_2-t_1) (f_{1,2} + f_{1,12}) + ( 1 \leftarrow\!\!\rightarrow 2),
\end{equation}
since $\frac{dt_{12}}{dt_2} = 1$ for $t_2 > t_1$, and $\frac{dt_{12}}{dt_1} = 1$ for $t_1 > t_2$. The two $f_{1,2}$ terms combine to form $\int_{\mathbb{R}_{+}^2} dt_1dt_2 f_{1,2}$, which is the term $R = \{1,2\}$ of (\ref{eqn10}). The other two integrals are the terms $R = \{1\}, \{2\}$. 

One can prove the general case by induction on $N$. Begin as above with
\begin{equation}\label{e32}
f({\bf 0}) = - \int^\infty_0 ds \sum^N_{k=1} f_k(s,\ldots,s,0,\ldots,0),
\end{equation}
where the integral is along the diagonal, $t_1 = t_2 = \cdots = t_N$. Consider one of these terms, say $k=N$, and apply (\ref{eqn10}) in the variables $\tilde{t}_i = t_i-t_N$, $i=1,\ldots,N-1$, keeping $t_N=s$ fixed:
\begin{equation}\label{e33}
 f_N(t_N,\ldots,t_N,0,\ldots,0) = \sum_{(\tilde{F},\tilde{R})} \int_{\mathbb{R}_{+}^{N-1}} \prod_{r \in \tilde{R}} [-d \tilde{t}_r] \prod_{ji \in \tilde{F}} [-d(\tilde{t}_j - \tilde{t}_i)] f_N^{(\tilde{F},\tilde{R})}({\bf t}).
\end{equation}
Note that when computing the derivative of $f_N$ with respect to $\tilde{t}_r$, there will be a term $f_{N,r}$ and also a term $f_{N,rN}$ (coming from the dependence on $t_{rN} = t_r-t_N = \tilde{t}_r$). Thus each $(\tilde{F},\tilde{R})$ on $\{1,\ldots,N-1\}$ gives rise to $2^{|\tilde{R}|}$ terms, each of which can be assigned a unique $(F,R)$ on $\{1,\ldots,N\}$. $R$ consists of $N$, together with each $r \in \tilde{R}$ with an $f_{N,r}$ term. $F$ consists of $\tilde{F}$, together with $rN$, $r \in \tilde{R}$ when $r$ gives rise to an $f_{N,rN}$ term. Observe that each root in $\tilde{R}$ ceases to be a root in $R$ if it is connected by a bond $rN$ in $F$. We obtain in this way all $(F,R)$ with $N \in R$, and each satisfies the condition that each tree of $F$ contains exactly one root. As a result,
\begin{equation}\label{last}
 f({\bf 0}) = \sum^N_{k=1} \sum_{(F,R): k \in R} \int_{\mathbb{R}^N_{+}} [-dt_k] \prod_{i \in R \setminus \{k\}} [-d(t_i-t_k)] \prod_{ji \in F}[-d(t_j-t_i)] f^{(F,R)}({\bf t}).
\end{equation}
It is evident that if we take the term $(F,R)$ of (\ref{eqn10}), and consider the subset of the integration region for which $t_k = \min_{r\in R} t_r$, we obtain the term $k,(F,R)$ of (\ref{last}). 

The forest-root formula (\ref{eqn10}) is the key to dimensional reduction; in fact a two-dimensional version of (\ref{eqn10}) was used in \cite{BI01,BI03} to derive a repulsive gas mapping for ordinary (isotropic) BP.  The argument proceeds by applying (\ref{eqn10}) to
\begin{equation}\label{eqn13}
f(\mathbf{t}) = g(t_1 / \epsilon) \prod^N_{i=2} g(\epsilon t_i)\prod_{1 \leq i < j \leq N} U(t_{ij},x_{ij}),
\end{equation}
where $g$ is any smooth function which decreases to 0 and satisfies $g(0) =1$. Thus,
\begin{align}\label{eqn14} 
 \rho_{\mathrm{HC}}(z) &= \lim_{\Lambda \nearrow S} \lim_{\epsilon \searrow 0} \frac{1}{Z_{\mathrm{HC}}(z)} \sum^\infty_{N=1} \frac{z^N}{N!} \int \prod^N_{i=1} dx_i f(\mathbf{0})\nonumber\\
 &= \lim_{\Lambda \nearrow S} \lim_{\epsilon \searrow 0} \frac{1}{Z_{\mathrm{HC}}(z)} \sum^\infty_{N=1} \frac{z^N}{N!} \sum_{(F,R)} \int_{(\mathbb{R}_{+}\times \Lambda)^{N}}
\prod_{r \in R} [-dy_{r} ] \prod_{ji \in F} [-dy_{ji}] f^{(F,R)}(\mathbf{t}) .
\end{align}
For each $ji \in F$, $U(y_{ji})$ is differentiated and becomes the linking weight $V(y_{ji})$. For each $r \in R$, a $g$ is differentiated: 
\begin{equation}
f^{(F,R)}(\mathbf{t})=\prod_{r\in R} g' \prod_{i \notin R} g \prod_{ji\in F} V(y_{ji}) \prod_{ji\notin F} U(y_{ji}).
\end{equation}
Note that the distribution of $t_r$ for an $n$-vertex tree is essentially $-(g(\epsilon t_r)^n)'dt_r$, which is a very spread-out probability measure.  One finds that due to the large separation in the $t$ direction, all the interaction terms $U$ between different trees are equal to 1 (up to terms which vanish as $\epsilon \rarrow 0$). This means that the sums/integrals associated with each tree are independent of the others, and all the trees of $F$ decouple.  One tree has its root fixed at 0 by a factor $-(g(t_1/\epsilon))'$, which converges to $\delta(t_1)$. The others cancel with the normalization $Z_{\mathrm{HC}}(z)$, so that (\ref{eqn14}) reduces to (\ref{eqn1}), with an additional $-1$ for each link of the tree.  Thus $\rho_{\mathrm{HC}}(z)=-Z_{\mathrm{DBP}}(-z)$, which is Theorem \ref{thm1}.

One can prove identities relating $n$-point functions of the repulsive gas to $n$-point functions of DBP. If one takes the derivative of (\ref{eqn5}) with respect to a source at $x$, one obtains for the 2-point correlation functions
\begin{equation}
G_{\mathrm{HC}}(0,x;z)=-\int_0^\infty G_{\mathrm{DBP}}(0,y;-z)dt,
\end{equation}
where $y=(t,x)\in \mathbb{R}_+ \times \mathbb{R}^D$.  Hence the transverse correlation length exponent $\nu_{\perp}(D+1)$ of the repulsive gas must be the same as $\nu(D)$. The value $\nu(1)=\frac{1}{2}$ can be computed directly, and the value $\nu(2)=\frac{5}{12}$ follows from hyperscaling ($D\nu=2-\alpha$) (or equivalently from $D\nu_{\perp}=2-\alpha$ \cite[Equation 28]{Fam82}) with $\alpha=\frac{7}{6}$ as indicated above. One can also consider unrooted DBP (divide each term in (\ref{eqn1}) by $N$) and relate them to the pressure of the associated repulsive gas.

In conclusion, we have demonstrated the underlying mechanism of dimensional reduction for directed branched polymers.  In the process, some lattice and continuum models of DBP are solved exactly in two and three dimensions by reference to repulsive gases in one lower dimension.
\section*{Acknowledgement}
 I thank David Brydges and John Cardy for conversations which improved my understanding of this problem. Research supported by NSF grant PHY-0244884.


\begin{thebibliography}{30}
 \setlength{\parskip}{0ex}%
      \setlength{\itemsep}{0ex}%


\bibitem{PS79} Parisi G and Sourlas N 1979 \textit{Phys. Rev. Lett.} {\bf 43} 744

\bibitem{I84} Imbrie J Z 1984 \textit{Phys. Rev. Lett.} {\bf 53} 1747

\bibitem{I85} Imbrie J Z 1985 {\it Commun. Math. Phys.} {\bf 98} 145

\bibitem{BDD98} Br{\'e}zin E and De~Dominicis C 1998 {\it Europhys. Lett.} {\bf 44} 13 
(\textit{Preprint} \href{http://arXiv.org/abs/cond-mat/9804266}{cond-mat/9804266})

\bibitem{F02} Feldman D E 2002 \textit{Phys. Rev. Lett.} {\bf 88} 177202 (\textit{Preprint} \href{http://arXiv.org/abs/cond-mat/0010012}{cond-mat/0010012})

\bibitem{PS02} Parisi G and Sourlas N 2002 \textit{Phys. Rev. Lett.} {\bf 89} 257204 (\textit{Preprint} \href{http://arXiv.org/abs/cond-mat/0207415}{cond-mat/0207415})

\bibitem{PS81} Parisi G and Sourlas N 1981 \textit{Phys. Rev. Lett.} {\bf 46} 871

\bibitem{BI01} Brydges D C and Imbrie J Z 2003 {\it Ann. Math.} {\bf 158} 1019
(\textit{Preprint} \href{http://arXiv.org/abs/math-ph/0107005}{math-ph/0107005})

\bibitem{BI03} Brydges D C and Imbrie J Z 2003 {\it J. Statist. Phys.} {\bf 110} 503
(\textit{Preprint} \href{http://arXiv.org/abs/math-ph/0203055}{math-ph/0203055})

\bibitem{F78} Fisher M E 1978 \textit{Phys. Rev. Lett.} {\bf 40} 1610

\bibitem{LF95} Lai S N and Fisher M E 1995 \textit{J. Chem. Phys.} {\bf 103} 8144

\bibitem{FP99} Park Y and Fisher M E 1999 \textit{Phys. Rev. } E {\bf 60} 6323 
(\textit{Preprint} \href{http://arXiv.org/abs/cond-mat/9907429}{cond-mat/9907429})

\bibitem{B} Baxter R J 1982 \textit{Exactly solved models in statistical mechanics} (London: Academic Press Inc.)

\bibitem{D83} Dhar D 1983 \textit{Phys. Rev. Lett.} {\bf 51} 853

\bibitem{BL87} Baram A and Luban M 1987 \textit{Phys. Rev. } A {\bf 36} 760

\bibitem{C85} Cardy J L 1985 \textit{Phys. Rev. Lett.} {\bf 54} 1354

\bibitem{C82} Cardy J L 1982 \textit{J. Phys. A: Math. Gen. } {\bf 15} L593

\bibitem{BJ82} Breuer N and Janssen H K 1982 \textit{Z. Phys.} B {\bf 48} 347

\bibitem{SRY82} Stanley H E, Redner S and Yang Z R 1982 \textit{J. Phys. A: Math. Gen. } {\bf 15} L569

\bibitem{DL82} Day A R and Lubensky T C 1982 \textit{J. Phys. A: Math. Gen. } {\bf 15} L285

\bibitem{RY82} Redner S and Yang Z R 1982 \textit{J. Phys. A: Math. Gen. } {\bf 15} L177

\bibitem{D82} Dhar D 1982 \textit{Phys. Rev. Lett.} {\bf 49} 959

\bibitem{C01} Chang S C and Shrock R 2001 {\it Physica} A {\bf 296} 131
(\textit{Preprint} \href{http://arXiv.org/abs/cond-mat/0005232}{cond-mat/0005232})

\bibitem{D03} Sumedha and Dhar D 2003 \textit{J. Phys. A: Math. Gen. } {\bf 36} 3701 (\textit{Preprint} \href{http://arXiv.org/abs/cond-mat/0303450}{cond-mat/0303450})


\bibitem{B98} Bousquet-M{\'e}lou M 1998 {\it Discrete Math.} {\bf 180} 73

\bibitem{RM01} van Rensburg E J J and Rechnitzer A 2001 {\it J. Statist. Phys.} {\bf 105} 49

\bibitem{K02} Kne{\v{z}}evi{\'c} M and Vannimenus J 2002 \textit{J. Phys. A: Math. Gen. } {\bf 35} 2725
(\textit{Preprint} \href{http://arXiv.org/abs/cond-mat/0203367}{cond-mat/0203367})


\bibitem{diF02} Di Francesco P and Guitter E 2002 \textit{J. Phys. A: Math. Gen. } {\bf 35} 897
(\textit{Preprint} \href{http://arXiv.org/abs/cond-mat/0104383}{cond-mat/0104383})

\bibitem{AR} Abdesselam A and Rivasseau V 1995 \textit{Constructive Physics} (\textit{Lecture Notes in Physics} vol 446) ed V Rivasseau (Berlin: Springer) p 7 (\textit{Preprint} \href{http://www.ma.utexas.edu/mp_arc-bin/mpa?yn=94-291}{mp\_arc:94-291})

\bibitem{Fam82} Family F 1982 \textit{J. Phys. A: Math. Gen. } {\bf 15} L583

\end{thebibliography}
\end{document}